\documentstyle[sprocl]{article}
\input{epsf}
\def\Journal#1#2#3#4{{#1} {\bf #2}, #3 (#4)}

\def\NPB{{\em Nucl. Phys.} B}

\def\PRL{\em Phys. Rev. Lett.}
\def\PRD{{\em Phys. Rev.} D}
\def\ZPC{{\em Z. Phys.} C}

\def\be{\begin{equation}}
\def\ee{\end{equation}}
\def\bea{\begin{eqnarray}}
\def\eea{\end{eqnarray}}

\begin{document}
\hfill{ANL-HEP-CP-97-41}
\vskip 0.1cm
\hfill{June 13, 1997}

\title{ISOLATED PROMPT PHOTON PLUS JET PHOTOPRODUCTION AT HERA
\footnote{Talk Presented at Photon'97, Egmond aan Zee, The Netherlands,
May 10-14, 1997} }

\author{ L. E. GORDON}

\address{Argonne National Laboratory, 9700 S. Cass Avenue,
Argonne,\\ IL 60439, USA}


\maketitle\abstracts{
The cross section for photoproduction of a prompt photon in association
with a jet is studied in Next-to-Leading Order at the DESY $ep$ collider
HERA. The effect of various cuts imposed on the cross section by the
ZEUS collaboration including isolation cuts on the photon is examined.
Comparisons with the ZEUS preliminary data using various
parametrizations of the photon structure function is made, and good
agreement is found. The preliminary data is not yet precise enough to make a
distinction between various models for the photon structure function.  
}

\section{Introduction}

It has long been anticipated that the DESY $ep$ collider HERA would
provide a good opportunity to study prompt photon production in
photoproduction processes \cite{aur1}. Over the past few years various
studies of this process have been performed with continuous improvements
in their theoretical precision \cite{bks,afg,gs,gv1}. In the most
recent studies \cite{gv1,gv2} the inclusive cross section for producing
a single photon was calculated fully in NLO with isolation effects
incorporated. In \cite{gv1,gv2} an approximate but nevertheless very
accurate analytic technique \cite{gv3,ggrv} for including isolation effects 
in the NLO calculation, including the fragmentation contributions was used. 
This analytic technique is only applicable to single inclusive prompt photon
production and cannot be applied when a jet is also observed.

The ZEUS Collaboration have begun analyzing prompt photon data
\cite{zeus} and have first chosen to look for events with a jet
balancing the transverse momentum ($p_T^{\gamma}$) of the photon. In order to
compare with this data a new calculation is necessary which will be
described in outline in the next section.

In all the previous studies of prompt photon production at HERA, one of
the common themes was the possibility of using it for measuring the 
photon distribution functions, particularly the gluon distribution,
$g^{\gamma}(x,Q^2)$ which is presently very poorly constrained by the available
data. This latter fact is still true even with the availability of jet 
photoproduction data at both HERA and TRISTAN. Prompt 
photon production is particularly attractive since it is dominated in Leading
Order (LO) by the hard scattering subprocess $q g\rightarrow \gamma q$,
resulting in a cross section which is very sensitive to the gluon
distribution. 

At HERA the situation is rather more complicated than at hadron
colliders by two factors. Firstly there are two particles involved in the
reaction, namely the quasi-real photon emitted by the electron which 
scatters at a small angle, as well as the proton. Both particles have
distinct gluon distribution functions $g^{\gamma}$ and $g^p$, hence two
different $qg$ initiated subprocess are present,
$q^p g^{\gamma}\rightarrow \gamma q$ and  $q^{\gamma} g^p\rightarrow
\gamma q$. Since they contribute to the cross section in different
regions of pseudo-rapidity, $\eta$, it has been proposed that this may provide 
a means of separating them experimentally, but this has proven to be very
difficult experimentally. Secondly, there are two types of contributions
to the cross section in photoproduction processes, usually labelled the direct
and resolved contributions. In the former case the quasi-real photon
participates directly in the hard scattering subprocess and gives up all
its energy, while in the latter, resolved, case it interacts via its
partonic substructure. Thus the resolved subprocesses are sensitive to
the photon structure functions whereas the direct are not. Again it was
proposed that they may be separated experimentally with suitable
rapidity cuts. A study performed in \cite{gs} has shown that since the
initial photon energy is not fixed but forms a continuous spectrum, then
even this separation is not straightforward. Separation of resolved and
direct processes is better achieved by tagging of the spectator jet from
the resolved photon.

\section{The Inclusive Photon Plus Jet Cross Section}


In addition to the direct and resolved photon contributions to the cross
section there are the non-fragmentation and fragmentation contributions.
In the former case the observed final state photon is produced directly in the 
hard scattering whereas in the latter it is produced by long distance
fragmentation off a final state parton. The fragmentation processes
involve the fragmentation functions which require a non-perturbative
input from experiment and have not been satisfactorily
measured up to now.

The only direct non-fragmentation process contributing to the cross section in 
LO is the so called QCD Compton process 
\begin{displaymath} 
\gamma q\rightarrow \gamma q
\end{displaymath}
and the direct fragmentation processes are
\begin{displaymath}
\gamma q\rightarrow g q\;\;\; {\rm and}\;\;\; \gamma g\rightarrow q\bar{q}.
\end{displaymath}
As discussed in many places (see eg. \cite{gv1}) the photon fragmentation
function is formally $O(\alpha_{em}/\alpha_s)$, thus although the hard
subprocess cross sections in the fragmentation case are
$O(\alpha_{em}\alpha_s)$, after convolution with the photon fragmentation
functions the process is $O(\alpha_{em}^2)$, the same as the the
non-fragmentation part.

At NLO for the non-fragmentation part there are the virtual corrections
to the LO Compton process plus the additional three-body processes
\begin{displaymath}
\gamma q\rightarrow \gamma g q\;\;\; {\rm and}\;\;\; \gamma g\rightarrow
\gamma q\bar{q}.
\end{displaymath}
In addition there are $O(\alpha_s)$ corrections to the fragmentation
processes to take into account, but in this calculation these processes
are included in LO only. It has been shown in \cite{gv1} that the
fragmentation contributions are not as significant here as at hadron
colliders which generally have higher cms energies and are reduced drastically 
when isolation cuts are implemented. Thus ignoring NLO corrections to the
fragmentation contributions while in principle inconsistent will not
lead to a large numerical error.

In the resolved case, for non-fragmentation there are only the two
processes  
\begin{displaymath}
 q g\rightarrow \gamma q\;\;\; {\rm and}\;\;\; q \bar{q}\rightarrow \gamma g.
\end{displaymath}
in LO. At NLO there are virtual and three-bofy corrections to these as well as 
other three-body processes, eg. $g g\rightarrow \gamma q \bar{q}$ etc. for a 
complete list of these plus the fragmentation processes see eg. \cite{gv1}.

The calculation was performed using the phase space slicing method which
makes it possible to perform photon isolation exactly as well as to
implement the jet definition. Details of the calculation can be found in
ref.\cite{gordon}. Following the ZEUS experiment, the cone
isolation method is used, which restricts the hadronic energy allowed in
a cone of radius $R_{\gamma}=\sqrt{\Delta \phi^2 + \Delta \eta^2}$,
centred on the photon to be
below the value $\epsilon E_{\gamma}$, where $E_{\gamma}$ is the photon
energy. The fixed value $\epsilon= 0.1$ is used which corresponds to the
value used in the ZEUS analysis. 

The cone algorithm is used to define the jet. This defines a jet as
hadronic energy deposited in a cone radius $R_J=
\sqrt{\Delta \phi^2 + \Delta \eta^2}$. If two partons form the jet then
the kinematic variables are combined to form that of the jet according to the
formulae 
\begin{eqnarray}
p_J&=&p_1+p_2 \nonumber \\
\eta_J&=& \frac{(\eta_1 p_1 + \eta_2 p_2)}{p_1+p_2} \nonumber \\
\phi_J&=& \frac{(\phi_1 p_1 + \phi_2 p_2)}{p_1+p_2}.
\end{eqnarray} 
In the ZEUS analysis $R_{\gamma}=1.0$ and $R_J=1.0$ are chosen and we
therefore use these values.

In order to estimate the flux of quasi-real photons from the electron
beam the Weiszacker-Williams approximation is used. Thus the `electron
structure function' $f_e(x_e,Q^2)$ is given by a convolution of the
photon structure function $f^{\gamma}(x_{\gamma},Q^2)$ and the
Weiszacker-Williams function
\begin{eqnarray}
f_{\gamma/e}(z)&=&\frac{\alpha_{em}}{2\pi}\left[\left\{
\frac{1+(1-z)^2}{z}\right\} \ln\frac{Q^2_{max}(1-z)}{m_e^2 z^2} 
\right. \nonumber \\
& - & \left. 2 m_e^2
z^2 \left\{ \frac{(1-z)}{m_e^2 z^2}-\frac{1}{Q^2_{max}}\right\} \right]
\end{eqnarray}
by
\begin{equation}
f_e(x_e,Q^2)=\int^1_{x_e}\frac{dz}{z}f_{\gamma/e}(z)f^{\gamma}\left(\frac{x_e}
{z},Q^2\right).
\end{equation}
The expression for $f_{\gamma/e}(z)$ was taken from ref.\cite{gas}.
Following the ZEUS analysis the value $Q^2_{max}=1$ GeV$^2$ is used
throughout. 

\section{Numerical Results}
The numerical results presented in this section are obtained using the
GS96 \cite{gs96} photon distribution functions, the CTEQ4M \cite{cteq} parton
distributions for the proton and the GRVLO \cite{grvf} fragmentation functions 
as standard. Futhermore the two-loop expression for $\alpha_s$ is used, 
four-flavours of quarks are assumed active and the
factorization/renormalization scales are taken to be equal to the photon
$p_T$ ($Q^2=(p_T^{\gamma})^2$). The maximum virtuality of the initial
state photon is fixed at $Q^2_{max}=1$ GeV$^2$ as chosen by the ZEUS
Collaboration. 
\begin{figure}
{\hskip 0.2cm}\hbox{\epsfxsize5.2cm\epsffile{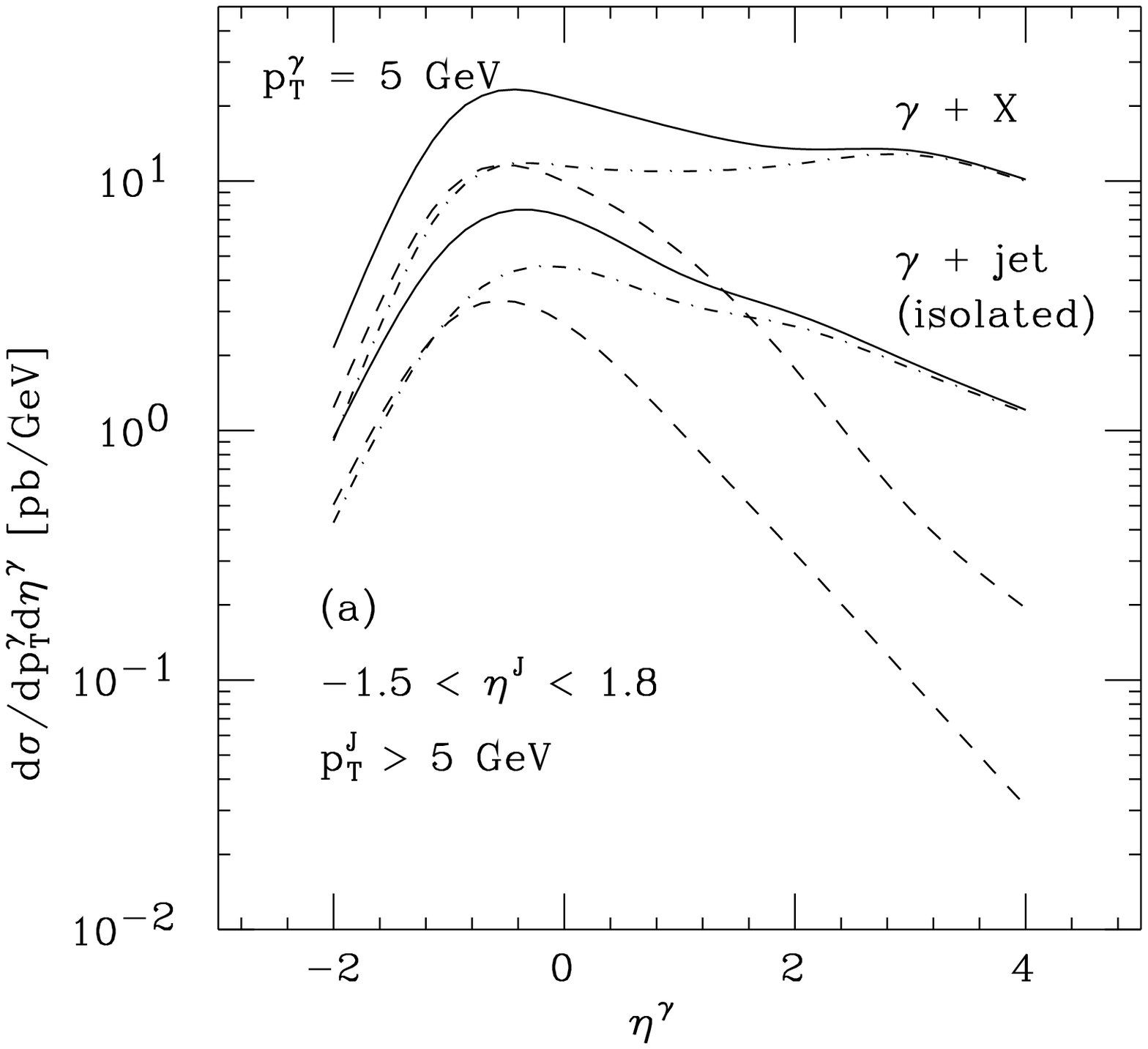}{\hskip 0.2cm}
\epsfxsize5.2cm\epsffile{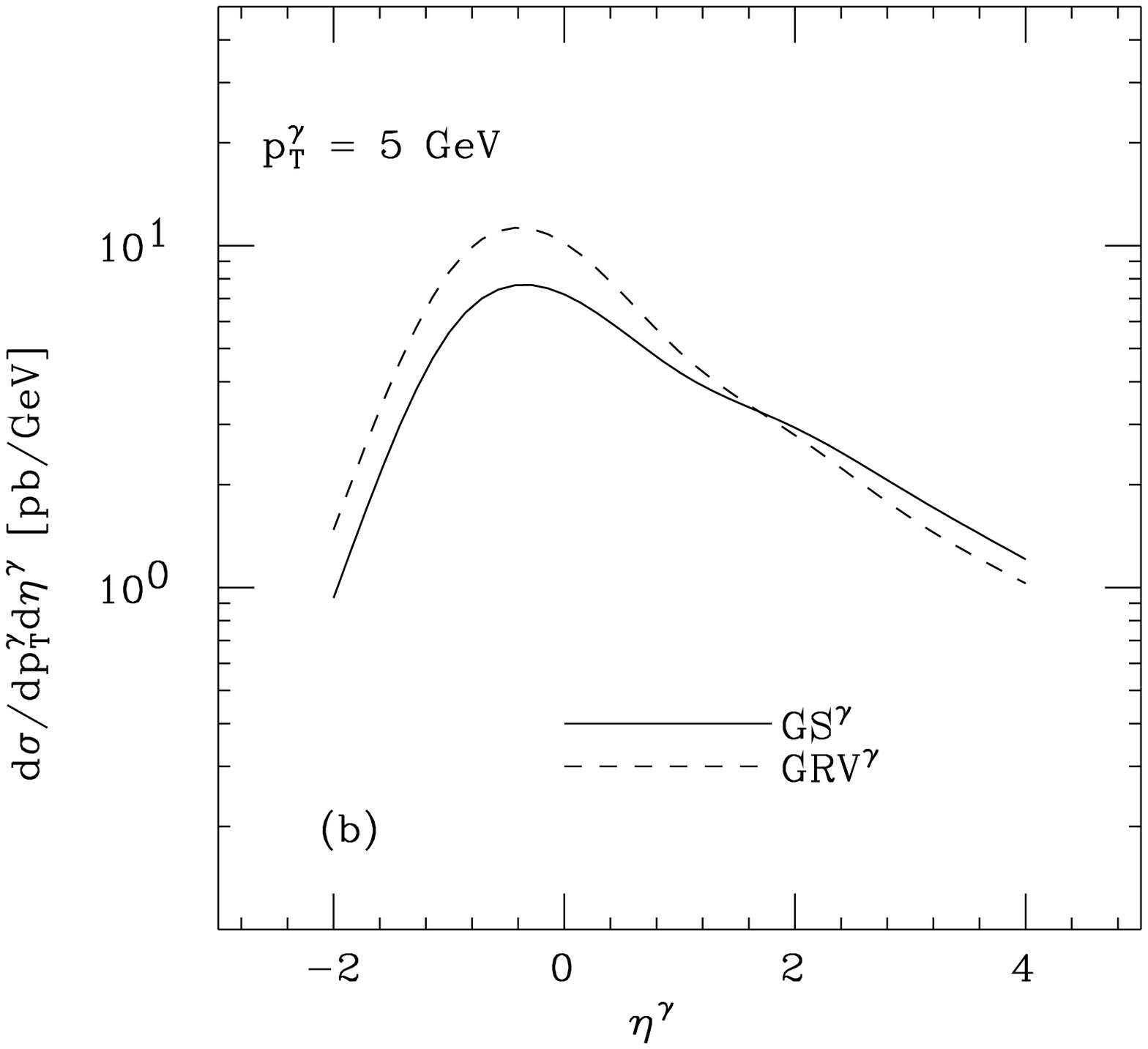}}
\caption{(a) Comparison of differential cross section for single $\gamma$ and
$\gamma+jet$. (b) Differential cross section with HERA (ZEUS) cuts for GRV and
GS96 photon distributions.} 
\end{figure}
Using the above parameters the fragmentation contribution constituted
less than $20\%$ of the cross section at $p_T^{\gamma}=5$ GeV before
isolation and falls rapidly with increasing $p_T^{\gamma}$. After
isolation this figure is reduced by $85\%$. The higher order corrections
enhances the cross section $O(20\%)$ before isolation. Imposing the ZEUS
$p_T$, rapidity, and isolation cuts on the inclusive single photon cross
section results in a drop in the cross section by more then a factor of
$3$ as shown in fig.1a . This severely reduces the accuracy of the 
measurement and makes it less likely that it will be useful in
distinguishing between the available photon distribution function
parametrizations. But as fig.1b shows if enough statistics can be
accumulated in the negative rapidity region the possibility still exists
to discriminate between the GRV \cite{grvg} and GS parametrizations.

Table 1 shows predictions for the resolved and direct contributions to
the cross section in $pb$ and their sum for various choices of parameters. In
order to obtain a sample of direct events the ZEUS Collaboration have
imposed the cut $x_{\gamma}\geq 0.8$ on their data. This cut which is
also imposed on the results in Table 1, favours the direct contributions
since they contribute at $x_{\gamma}=1$, but there is still a
contribution from the resolved processes and hence some sensitivity to
the photon distributions chosen. In addition the cuts $5$ GeV
$\leq p_T^{\gamma} \leq 10$ GeV, $p_T^J\geq 5$ GeV, $-1.5\leq \eta^J\leq
1.8$, $-0.7\leq \eta^{\gamma}\leq 0.8$ and $0.16\leq z=E_{\gamma}/E_e\leq 0.8$ 
along with the isolation
cuts and jet definitions discussed in section 2 are imposed. 

The first column of numbers gives the results for the standard choice of 
parameters, while the 2nd and 3rd columns show the effect of changing the 
scales. The results
show a remarkable stability to scale changes. This is in contrast to eg.
the $p_T^{\gamma}$ distribution which generally shows significant scale
sensitivity. The 4th and 5th columns show the effect of changing the photon
and proton distribution functions used respectively. In the latter case
there is hardly any changes in the predictions, while in the former case
the changes are very significant. Since with these cuts the cross
section is mostly sensitive to the quark distributions in the photon at
large-$x$ then it may potentially be used to discriminate between the
GS96 and GRV photon parametrizations which differ most significantly in
this region. The preliminary experimental value given by the ZUES
Collaboration of $17.1\pm 4.5\pm 1.5$ pb agrees remarkably well with the NLO
theoretical predictions but the errors quoted are still too large to make any
distinction between GS and GRV.

\begin{table}[t]
\caption{Total $\gamma + jet$ cross section with HERA cuts (see text).
\label{tab:exp}}
\vspace{0.4cm}
\begin{center}
\begin{tabular}{|r|r|r|r|r|l|}
\hline   
 &STD&$Q^2=(p_T^{\gamma})^2/4$&$Q^2=4(p_T^{\gamma})^2$&GRV$^{\gamma}$
                                                             &MRSR1\\
\hline
RES&3.31  &2.60 &4.95 &6.72 &3.44   \\
\hline
DIR&9.86 &11.45 &8.18 &9.86 &9.34 \\
\hline
SUM &$13.17$ &$14.05$ &$13.13$ &$16.58$ &$12.78$ \\
\hline
\end{tabular}
\end{center} 
\end{table}

\section{Conclusions}

A NLO calculation of isolated single photon plus jet production at HERA was
presented and compared to the preliminary data from the ZEUS
Collaboration and good agreement was found. The kinematic cuts chosen
favour the direct contribution but there is still a significant
sensitivity to the quarks distributions in the photon at large-$x$. At
the moment the error in the data is still too large to distinguish
between the GRV and GS96 photon distributions, but it is expected that
accumulation of more data will soon remedy this situation.  

\section*{Acknowledgments}
I am indebted to E. L. Berger for helping me to secure funds to
attend the Photon'97 conference and to the Planning Committee for
waiving the registration fee. This work was funded in part by the US
Department of Energy, Division of High Energy Physics, Contract No.
W-31-109-ENG-38.

\end{document}